\documentclass{article}

\usepackage[dblblindworkshop, final]{neurips_2026}
\workshoptitle{On-Device Intelligence: Foundation Models under Real-World Constraints}

\usepackage[utf8]{inputenc}
\usepackage[T1]{fontenc}
\usepackage{hyperref}
\usepackage{url}
\usepackage{xurl}
\usepackage{booktabs}
\usepackage{array}
\usepackage{enumitem}
\usepackage{amsmath}
\usepackage{graphicx}
\usepackage{tikz}
\usepackage{pgfplots}
\pgfplotsset{compat=1.17}
\usetikzlibrary{arrows.meta,positioning,fit,calc,backgrounds}
\usepackage[capitalise,noabbrev]{cleveref}
\usepackage{placeins}

\makeatletter
\renewcommand{\@notice}{%
  \enlargethispage{2\baselineskip}%
  \@float{noticebox}[b]%
    \footnotesize NeurIPS 2026 Workshop on On-Device Intelligence: Foundation Models under
    Real-World Constraints.%
  \end@float%
}
\makeatother

\newcommand{\app}{\texttt{omni-macos}}
\newcommand{\modelfamily}{\texttt{jina-embeddings-v5-omni}}
\newcommand{\modelnano}{\texttt{jina-embeddings-v5-omni-nano}}
\newcommand{\modelsmall}{\texttt{jina-embeddings-v5-omni-small}}

\title{\app{}: On-Device Omni-Modal Search on Apple Silicon}

\author{%
  Han Xiao \\
  Jina AI \textit{by} Elastic \\
  \texttt{han.xiao@jina.ai} \\
}

\begin{document}

\maketitle

\begin{abstract}
We present \app{}, a search engine that embeds text, code, documents, images, audio and video into
one representation space and runs on the Mac that already holds the files, so the files, the queries
and the vectors stay on that machine. A background indexer and an interactive search box share the
device. Each distinct chunk is
embedded and stored once, and every file becomes searchable from a small first share of its
content. Indexer work is split into smaller units while the user types, and one memory setting
reaches every allocator found to draw on unified memory. Once the index is large, queries scan a
one-bit replica of it, rescore a shortlist exactly and rank every modality in one list on a common score scale. The
engine can also leave the GPU entirely, with the encoder on the Neural Engine and search on the
CPU. We measure \app{} on five Macs spanning an eightfold range of accelerator width and a thirty-twofold
range of memory, and its ranking on a labeled multilingual and multimodal testbed. \app{} is open
source under Apache 2.0 at \url{https://github.com/hanxiao/omni-macos}.
\end{abstract}

\section{Introduction}

Semantic search over personal files is assembled from parts that are individually solved. One
multimodal encoder can embed text, code, documents, images, audio and video into the same
representation space. Quantization has made storing the vectors cheap, and any vector database will
hold the result at the scale one person reaches. Almost every one of those parts assumes a server,
so the files are uploaded, embedded on hardware belonging to someone else, and searched there. For
the files worth keeping private, that is often disqualifying.

\app{} runs the whole engine in one process on the machine that already holds the files, so the
files, the queries and the vectors never leave it. That decision moves the difficulty from the model to the system: the work a server would absorb now
runs inside a multitasking operating system, beside everything else the user is running. The
hard part is keeping an index current
with files the user is actively editing while a search box stays quick. The files appear, change and
are deleted while the system runs, so the index is never finished, and the cost that matters most
is the marginal cost of saving one file, most of whose text the index already holds. A first pass
over a large library still takes hours, and a file it has not reached cannot be found. A removed folder or a disabled file type changes much of the index at once. An
indexer that takes over the device to finish sooner serves the user worse than one that takes longer
and goes unnoticed. On Apple silicon the memory is unified:
the GPU has no dedicated memory and allocates from the same system memory as the
CPU~\citep{apple2026storagemodes,apple2026devicememory}. What the encoder and the resident index
take is therefore taken from the applications the user is working in, and the system responds to
exhaustion by compressing memory and swapping.

We contribute the following.

\begin{itemize}[leftmargin=1.2em,itemsep=1pt,topsep=3pt]
\item The design and implementation of an omni-modal search engine that runs entirely inside one
process on consumer Apple silicon. It addresses every chunk by its content so that each is embedded
and stored once, and makes every file searchable from its first part. It bounds what a continuous
indexer and a bulk change to the index can cost an interactive search, ranks every modality in one
list on a common score scale, and propagates the memory headroom the user sets to every allocator it
finds. It can also run every forward pass on the Neural Engine and every scan on the CPU, which
leaves the GPU to the other applications.
\item An account of what holding that headroom takes on unified memory, where CPU, GPU and framework
caches draw on one pool: the allocators the setting must reach cannot be listed up front, and one
number per allocator is not always enough.
\end{itemize}

\Cref{sec:device} describes what the system does, \Cref{sec:design} how it is built, and
\Cref{sec:eval} what each part costs and what the search retrieves, with supporting measurements
in the appendix.

\section{The system}
\label{sec:device}

\begin{figure}[!htbp]
\centering
\input{figures/appshot.tex}
\caption{\app{} answering \texttt{porsche sports car} within one folder of a real collection of
local files. One query returns one ranked list holding a PDF, located to its page, and photographs,
because every file type shares one representation space.}
\label{fig:app}
\end{figure}

\Cref{fig:app} shows the interface. \app{} is a native macOS application
for Apple silicon, written entirely in Swift over a Metal-backed array framework. Its interaction follows Finder conventions
throughout, so a result can be previewed, revealed in Finder or dragged out like any other file. The encoder is \modelfamily{}~\citep{akram2026jinav5text,honicke2026jinav5omni}, which ships in two sizes the user can
switch between, \modelnano{} at 0.95\,B parameters and \modelsmall{} at 1.57\,B. One encoder covers every
modality, so a text query can be scored against an image, an audio file or a video. It runs on
the GPU, or, if the user chooses, on the Neural Engine~\citep{orhon2022ane}, the separate
accelerator for neural-network inference on Apple silicon.

The crawler accepts the file types a working directory contains: plain text and markup, source
code, word-processing documents, presentations, spreadsheets, PDFs, and the common image, audio and
video containers. The user can add the Apple Photos library as a further source. Text and code are embedded as text, images as pixels, audio from its mel
spectrogram, and video by sampling frames, with its soundtrack embedded as audio. The text in a
screenshot or a scanned page is recognized on the device and indexed as well. Documents are routed
by their content: a PDF with a text layer is embedded as text, and a scanned one is rendered to page
images and embedded as images. A slide is embedded both by its words and as a picture, and so is a
page of a text PDF that is mostly an image, such as a photograph or a scanned insert. Large files are
cut into chunks that are embedded independently, and a hit opens at the line, page, slide, rows of a
sheet or moment that matched.
Indexing starts on the first files the crawler finds, and a long file is searchable once its first
part is indexed, so a large library becomes searchable while it is still being scanned. A file-system
watcher keeps the index current while the application runs.

The interface offers a search box that answers as the user types, a pivot from any result to others
like it, open-vocabulary tags on images and video,\footnote{Tags score vocabulary embeddings
against the patches the image forward pass already produced, as described at
\url{https://hanxiao.io/ttc-embedding-image-tagging-2026/}.} and a two-dimensional map of a folder,
laid out by PCA over the stored vectors and optionally refined by a UMAP-style force
layout~\citep{mcinnes2018umap}. Each result shows the passage of its file that is closest to the
query. Results can be restricted to a modality, a folder, a file extension, a tag or a recent period, and weaker matches are folded under one control. A file can also be found by its name or by
an identifier it contains, such as an invoice number. Several Macs running \app{} can be linked for
federated search: a query is answered by each linked Mac from its own index, and every index stays on
the Mac that holds its files. \app{} also serves other agents on the same machine over the Model
Context Protocol, with the search and the filters of the search box, and a command line reaches the
same tools.

\section{Design and implementation}
\label{sec:design}

\Cref{fig:arch} lays out the two paths that reach the encoder. The index path walks the indexed
folders, where the size and modification time of an entry decide whether it is already indexed. It
then extracts content, cuts it into chunks and hashes them, and encodes only the chunks whose content
the store does not already hold. The query path encodes the query, scans the store and returns. Both reach the accelerator through a single command queue, so their submissions execute
one after another and the two paths are never on the device at once. A priority gate decides which
of them submits next.

\begin{figure}[!htbp]
\centering
\pgfdeclarelayer{panels}\pgfdeclarelayer{blocks}\pgfdeclarelayer{links}
\pgfsetlayers{panels,blocks,links,main}
\definecolor{acc}{RGB}{31,84,140}
\definecolor{gpu}{RGB}{176,96,40}
\begin{tikzpicture}[
  panel/.style={fill=black!5, rounded corners=2.5pt, draw=none, inner sep=0.22cm},
  blk/.style={rounded corners=2pt, line width=0.4pt, minimum width=2.0cm, minimum height=0.6cm,
              font=\scriptsize, align=center, inner sep=1pt},
  qb/.style={blk, fill=acc!10, draw=acc!60},
  ib/.style={blk, fill=black!12, draw=black!45},
  gb/.style={blk, fill=black!22, draw=black!60},
  wb/.style={blk, fill=gpu!15, draw=gpu!75},
  ptitle/.style={font=\scriptsize, anchor=north west, inner sep=0pt},
  lab/.style={font=\tiny, inner sep=1.5pt},
  ln/.style={line width=0.5pt, draw=black!65},
  ar/.style={ln, -{Latex[length=1.3mm,width=1mm]}},
  qln/.style={ln, draw=acc!80},
  qar/.style={ar, draw=acc!80},
  sig/.style={qar, dash pattern=on 2pt off 1.4pt},
]
\def\cA{0}\def\cB{2.75}\def\cC{5.5}\def\cD{8.25}\def\cE{11}

\begin{pgfonlayer}{blocks}
  \node[qb] (agent) at (\cB,3.9) {dropped file\\or agent};
  \node[qb] (sbox)  at (\cA,3.0) {search box};
  \node[qb] (deb)   at (\cB,3.0) {debounce};
  \node[ib] (crawl) at (\cA,0)    {crawl};
  \node[ib] (watch) at (\cA,-0.9) {file watcher};
  \node[ib] (chunk) at (\cB,0)    {chunk};
  \node[gb] (gate)  at (\cC,1.5)  {priority gate};
  \node[wb] (enc)   at (\cD,1.5)  {encoder};
  \node[wb] (scan)  at (\cE,1.5)  {scan};
  \node[ib] (delta) at (\cD,0)    {delta};
  \node[ib] (base)  at (\cE,0)    {base};
\end{pgfonlayer}

\begin{pgfonlayer}{panels}
  \coordinate (tq) at ([yshift=0.32cm]agent.north);
  \coordinate (ti) at ([yshift=0.32cm]crawl.north);
  \coordinate (ts) at ([yshift=0.32cm]delta.north);
  \node[panel, fit=(agent)(sbox)(tq)] (pq) {};
  \node[panel, fit=(crawl)(watch)(chunk)(ti)] (pix) {};
  \node[panel, fit=(delta)(base)(ts)] (ps) {};
\end{pgfonlayer}
\node[ptitle] at ([xshift=0.22cm,yshift=-0.16cm]pq.north west) {Query path};
\node[ptitle] at ([xshift=0.22cm,yshift=-0.16cm]pix.north west) {Index path};
\node[ptitle] at ([xshift=0.22cm,yshift=-0.16cm]ps.north west) {Store};
\node[lab, anchor=south] at ([yshift=0.04cm]enc.north) {GPU or Neural Engine};
\node[lab, anchor=south] at ([yshift=0.04cm]scan.north) {GPU or CPU};

\begin{pgfonlayer}{links}
  \draw[qar] (sbox) -- (deb);
  \draw[qar] (deb.east) -- node[lab, above, pos=0.5] {query} (\cC,3.0) -- (gate.north);
  \draw[qln] (agent.east) -- (\cC,3.9) -- (\cC,3.0);
  \draw[ar] (crawl) -- (chunk);
  \draw[ln] (watch.east) -- (\cA+1.375,-0.9) -- (\cA+1.375,0);
  \draw[ar] (chunk.east) -- node[lab, above, pos=0.5] {new chunks} (\cC,0) -- (gate.south);
  \draw[sig] (deb.south) -- node[lab, left, pos=0.5] {typing} (\cB,1.5) -- (gate.west);
  \draw[ar] (gate) -- (enc);
  \draw[qar] (enc) -- node[lab, above] {query} (scan);
  \draw[ar] (enc.south) -- node[lab, right] {new vectors} (delta.north);
  \draw[ar] (delta) -- node[lab, above] {fold} (base);
  \draw[ar] (base.north) -- node[lab, right] {read} (scan.south);
\end{pgfonlayer}
\end{tikzpicture}
\caption{The query and index paths share one encoder behind a priority gate. While the user types (dashed), the gate admits indexer work in smaller units. Warm blocks run
on the first unit named above them, or on the second when the user chooses the Neural Engine. The scan reads the store without
passing the gate, and the fold moves new vectors from the delta into the base.}
\label{fig:arch}
\end{figure}

\subsection{The encoder}
\label{sec:arith}

Every file the crawler accepts is chunked and embedded, which makes the encoder the dominant cost in
the system. The encoder is our own implementation of both sizes of \modelfamily{}. The released checkpoints include task-specific LoRA adapters alongside
the base weights, and \app{} ships only the retrieval adapter, merged into the base once,
$W \mathrel{+}= (\alpha/r)\,BA$. Loading therefore costs no merge, and at runtime there is a single
dense set of weights. The output is a unit vector pooled from the last token, 768-dimensional for
\modelnano{} and 1024-dimensional for \modelsmall{}.

The implementation departs from a direct translation in three places: precision is chosen per
tensor, kernel launches are merged, and the final layer is narrowed to the rows that survive
pooling. These change floating-point rounding and nothing else, so our implementation reproduces the
vectors of the code released with the model up to rounding, which we check by cosine similarity.

Precision is chosen per tensor. The matrix multiplications and attention of the backbone run in bf16, which halves
the resident weights of the backbone. The vision tower stores its weights and computes in
fp16, which is as wide as bf16 and keeps three more bits of mantissa, so its vectors stay closer to
those of an fp32 pass. Its fused attention accumulates in fp32 inside the kernel, and a pass whose
output is not finite is computed again in fp32.
Attention dominates the time in that tower and is bound by the memory bandwidth of its operands, so
their width matters more than the format of everything around them.

A kernel launch is one call handed to the GPU, and one item does not come close to filling the
device. An image is roughly a thousand patches, well short of the sequence length at which fused
attention saturates, and a text chunk is shorter still. We therefore merge launches wherever the
shapes allow, since a launch that could have been merged leaves the device idle between kernels.
The vision tower
is the clearest case, where a per-window loop issues one attention call per window. When the windows
share one size, as in the frames of a video or a batch of same-size images, they are assembled into
a single block-diagonal call. The images of one call then pass through the backbone together,
grouped by length so that padding stays small. Inside the attention of that tower, the rotary
embedding, the casts and the change of layout that fused attention expects are one kernel of our own. The same
pattern turns a slice-and-stack over the batch in the backbone into one gather operation, and lets
the pooled rows of a batch be evaluated in one call.

Tail-row narrowing shortens the final layer. The embedding is pooled from the last token, so
exactly one row per sequence survives, and attention is the last operation that mixes rows. Everything in the final layer after
attention, the post-attention norm, the MLP, the residual and the final norm, can therefore be
computed on the pooled rows alone. The saving equals the share of the
backbone that the final-layer MLP accounts for, which is 6.25\% for \modelnano{}, and no term is
dropped.

\subsection{Content addressing}
\label{sec:reuse}

The index identifies a chunk by what it contains. It
keeps one vector per distinct content, and a file is a sequence of references to contents. The
vector, snippet and kind of a chunk are therefore stored once, apart from the file, ordinal and
positional locator of each occurrence. Three parts of the system follow from
this separation: reuse, which becomes a lookup by content; the scan, which scores contents and
expands them to files; and deletion, which removes references. \Cref{fig:content} shows the separation on two files.

\begin{figure}[!htbp]
\centering
%
%
\pgfdeclarelayer{z2}\pgfdeclarelayer{z3}\pgfdeclarelayer{z4}
\pgfsetlayers{z2,z3,z4,main}
\definecolor{gpu}{RGB}{176,96,40}
\begin{tikzpicture}[
  x=0.195cm, y=0.195cm,
  occ/.style={fill=white, draw=black!45, line width=0.4pt, rounded corners=1.5pt},
  con/.style={fill=black!12, draw=black!45, line width=0.4pt, rounded corners=1.5pt},
  new/.style={fill=gpu!16, draw=gpu!75, line width=0.55pt, rounded corners=1.5pt},
  dead/.style={fill=black!3, draw=black!35, dashed, line width=0.4pt, rounded corners=1.5pt},
  box/.style={minimum width=1.56cm, minimum height=0.585cm, inner sep=0pt},
  lbl/.style={font=\scriptsize, inner sep=0pt, text height=1.45ex, text depth=0.35ex},
  row/.style={font=\scriptsize, anchor=west, inner sep=0pt},
  ann/.style={font=\tiny, align=center, inner sep=1pt},
  ar/.style={-{Latex[length=1.1mm,width=0.85mm]}, line width=0.5pt, draw=black!65},
]
\begin{pgfonlayer}{z2}
  \node[occ, box] (a1) at (16,14) {};
  \node[occ, box] (a2) at (26,14) {};
  \node[occ, box] (a3) at (36,14) {};
  \node[occ, box] (a4) at (46,14) {};
  \node[con,  box] (c1) at (16,7) {};
  \node[con,  box] (c2) at (26,7) {};
  \node[new,  box] (c3) at (36,7) {};
  \node[con,  box] (c4) at (46,7) {};
  \node[con,  box] (c5) at (56,7) {};
  \node[dead, box] (c6) at (66,7) {};
  \node[occ, box] (b1) at (46,0) {};
  \node[occ, box] (b2) at (56,0) {};
\end{pgfonlayer}

\begin{pgfonlayer}{z3}
  \foreach \i in {1,2,3,4} \draw[ar] (a\i.south) -- (c\i.north);
  \draw[ar] (b1.north) -- (c4.south);
  \draw[ar] (b2.north) -- (c5.south);
\end{pgfonlayer}

\begin{pgfonlayer}{z4}
  \node[row] at (0,14) {file A};
  \node[row] at (0,7)  {contents};
  \node[row] at (0,0)  {file B};

  \node[lbl] at (a1) {$p$};  \node[lbl] at (a2) {$q$};
  \node[lbl] at (a3) {$r'$}; \node[lbl] at (a4) {$s$};
  \node[lbl] at (b1) {$s$};  \node[lbl] at (b2) {$t$};
  \node[lbl] at (c1) {$p$};  \node[lbl] at (c2) {$q$};
  \node[lbl] at (c3) {$r'$}; \node[lbl] at (c4) {$s$};
  \node[lbl] at (c5) {$t$};  \node[lbl] at (c6) {$r$};

  \node[ann, anchor=south] at ([yshift=1pt]a3.north) {edited};
  \node[ann, anchor=north] at ([yshift=-2pt]c3.south) {encoded};
  \node[ann, anchor=south] at ([yshift=2pt]c6.north) {released};
\end{pgfonlayer}
\end{tikzpicture}
\caption{Content addressing on two files. The edit to the third chunk of file A adds the one content
that is encoded, the passage $s$ in both files is stored once, and the replaced content $r$ has no
references left, so its position is released.}
\label{fig:content}
\end{figure}

Reuse matters because keeping up with the files being edited is the dominant ongoing cost, and
most of that cost would re-derive vectors that already exist. A saved file has rarely changed
by much, and the same passage recurs across files as a shared header, a quoted file or a repeated
template. Both kinds of repetition skip the encoder once a chunk is identified by its content, but
only if the content of a chunk survives an edit to its neighbors.

Chunk boundaries are therefore content-defined~\citep{muthitacharoen2001,xia2016}. A gear hash
rolls over the bytes and declares a boundary where its low bits are zero. The mask is stricter
before the target size and looser after it, so sizes gather near the target, and each cut moves to
the next line break within a short window. Sizes are counted in characters, so a chunk
holds the same amount of text in every script, and the target is the chunk size the user sets. A
boundary depends only on the bytes around it. An edit therefore changes only the chunks around it, and a passage that occurs in two files becomes the
same chunk in both.

The key of a text chunk is a hash of its text together with the parameters that determine what it
means: those of the chunker and the embedding dimension. Changing any of them changes the key, so a
vector left stale by such a change cannot be silently reused. A picture or an audio window is keyed
by what the encoder receives, its preprocessed patches or its mel spectrogram, so an edit to the
metadata of a file leaves its key unchanged. A chunk reaches the encoder only if its key is found in
none of four places. The previous chunks of the same path answer an edited file with one query. A
bounded cache holds the vectors computed earlier in the same pass, and the content index covers the
whole store. A bounded pool keeps the vectors the store
released most recently, so a file restored from the trash or an edit undone is not encoded again. A
hit adds a reference and no vector. The locator is computed for each occurrence, because the same paragraph
can be line 12 of one file and line 4310 of another, and a line number moves whenever earlier text
changes length. A file whose bytes are identical to a file already indexed does not reach the
chunker at all. Its key is a digest over the bytes and the settings that decide what a vector
means, and a hit points the new file at the contents of the earlier one.

Content other than plain text is cut so that each chunk has a place a hit can open at. Audio is
cut into one-minute windows, and video into segments of up to four minutes, each embedded from
frames sampled across it, with the soundtrack cut and embedded as audio. A chunk of a PDF is located
at the page holding most of its text. A short PDF whose text fits in one chunk but spans several
pages is cut at its page breaks, because a single vector over unrelated pages ranks the document low
for a question about any one of them. A slide is one chunk, located by its number, and a sheet is
cut by rows into pieces of a quarter of a chunk, each opening with the name of the sheet and its header row and
located at the cells it holds. The picture of a slide, or of a PDF page with no text of its own or
with images over 30\% of its area, is a chunk at the same place; the share of images is read
from the drawing commands of the page, without rendering it. Text read from a screenshot or a
scanned page is chunked as text and stored beside the vector of the image.

The snippet the store keeps is only the opening of a chunk, while a result shows the passage that
matched. The text of a chunk is therefore recovered from its file when a result needs it: the
opening is found in the file, and the chunker, started there, cuts the same chunk again, because a
boundary depends only on the bytes around it. The key confirms the match, so a file changed since it
was indexed yields no passage. The recovered chunk is cut into windows, which are embedded at query
time for the results in rank order within a fixed time budget, and the window closest to the query
is shown.

\subsection{Coverage-first indexing}
\label{sec:turns}

The index path orders the first pass so that every file becomes searchable early. The walk takes
the most recently modified folder next and lists the files of each folder newest first, and a long
file is indexed in turns. Its first turn
indexes the least that makes it searchable, 64\,KB of text, one video segment, one minute of audio
or a first group of scanned pages, and its remainder goes to the end of the queue as a cursor. Later
turns are full size, chosen so that a turn of every kind costs the encoder about the same time, so
the total work is unchanged and only its order differs.

The cursors are kept beside the index, so a pass stopped between turns resumes where it stopped
and embeds nothing twice. The digest that lets an identical copy skip the chunker is recorded only
with the last turn, so a copy is never matched to a file that is still partly indexed. A file the
index already holds is not split into turns when it changes, since its unchanged chunks are found by
content in a single pass. The pass also yields to the state of the machine: it waits in Low Power
Mode and while the machine runs hot and, if the user chooses, on battery power or outside set
hours, and the changes the watcher reports are still indexed.

\subsection{Anticipatory shaping}
\label{sec:shape}

Once a command buffer is committed it runs to completion~\citep{apple2026commandbuffer}: the
interface offers no way to cancel it, and none to lower the priority of work already queued, so
priority can be expressed only at admission. This is ordinary for a GPU: preemption, where it
exists, is an architectural facility the driver uses to schedule between
processes~\citep{tanasic2014,park2015chimera}, and an application cannot apply it to its
own submissions. The gate is therefore a two-class non-preemptive lock: a waiting query is admitted
ahead of indexer work that has not started, and behind the work already in flight. Three moments
are early enough to act on: admission, the points inside a hold where nothing is in flight, and the
interval before a query is issued.

The second moment comes inside a hold, which is a sequence of submissions: the indexer embeds
several text batches under one hold, and an image call runs the vision tower chunk by chunk before
the backbone. Between two submissions nothing
of the hold is in flight, and at each such point the holder checks for a waiting query, hands it the
gate and takes the gate back once the query is done. A query therefore waits only for the work
already in flight, as if it had arrived between two holds, and the vectors the indexer writes are
unchanged. This serves a query from an agent, which has no keystroke before it, as well as one from
the search box. The engine also keeps the vectors of recent queries, so a repeated query, such as an
agent paging through results, skips the encoder.

\begin{figure}[!htbp]
\centering
%
%
%
%
\definecolor{acc}{RGB}{31,84,140}
\begin{tikzpicture}[
  x=2.44cm, y=1cm,
  lane/.style={font=\scriptsize, text=black, anchor=south west, inner sep=0pt},
  lbl/.style={font=\tiny, text=black, inner sep=1pt},
  acclbl/.style={font=\tiny, text=acc, inner sep=1pt},
  rule/.style={draw=black!22, line width=0.4pt},
  guide/.style={dashed, dash pattern=on 1.1pt off 1.1pt, draw=acc!70, line width=0.5pt},
  kguide/.style={dashed, dash pattern=on 1.1pt off 1.1pt, draw=black!45, line width=0.5pt},
  endguide/.style={dotted, draw=black!60, line width=0.6pt},
  wait/.style={-{Latex[length=1.2mm,width=0.95mm]}, draw=acc!85, line width=0.55pt},
  span/.style={{Latex[length=1mm,width=0.8mm]}-{Latex[length=1mm,width=0.8mm]},
               draw=black!50, line width=0.4pt},
]
\def\idxfill{black!13}
\def\idxdraw{black!45}

\draw[endguide] (5.62,0.55) -- (5.62,3.13);

\node[lane] at (0,3.14) {without shaping};
\draw[rule] (0,2.60) -- (5.62,2.60);
\foreach \a/\b in {0/1.2, 1.3/2.5, 2.6/3.8, 3.9/5.1}
  \filldraw[fill=\idxfill, draw=\idxdraw, line width=0.35pt] (\a,2.60) rectangle (\b,3.06);
\filldraw[fill=acc!30, draw=acc!80, line width=0.4pt] (5.2,2.60) rectangle (5.62,3.06);
\node[lbl] at (5.41,2.83) {query};

\draw[guide] (4.3,3.72) -- (4.3,3.06);
\node[lbl, anchor=south] at (4.3,3.76) {query arrives};

\draw[wait] (4.3,2.44) -- (5.18,2.44);
\node[lbl, anchor=north] at (4.74,2.36) {wait};

\node[lane] at (0,1.16) {with shaping};
\draw[rule] (0,0.62) -- (5.62,0.62);
\foreach \a/\b in {0/1.2, 1.3/2.5, 2.6/3.8}
  \filldraw[fill=\idxfill, draw=\idxdraw, line width=0.35pt] (\a,0.62) rectangle (\b,1.08);
\foreach \a/\b in {3.9/4.14, 4.16/4.40, 4.86/5.10, 5.12/5.36, 5.38/5.62}
  \filldraw[fill=\idxfill, draw=\idxdraw, line width=0.35pt] (\a,0.62) rectangle (\b,1.08);
\filldraw[fill=acc!30, draw=acc!80, line width=0.4pt] (4.42,0.62) rectangle (4.84,1.08);
\node[lbl] at (4.63,0.85) {query};

\draw[kguide] (3.5,1.72) -- (3.5,1.08);
\draw[guide]  (4.3,1.72) -- (4.3,1.08);
\node[lbl, anchor=south] at (3.5,1.76) {keystroke};
\node[lbl, anchor=south] at (4.3,1.76) {query arrives};
\draw[span] (3.5,1.46) -- (4.3,1.46);
\node[lbl, anchor=north] at (3.9,1.40) {debounce};

\draw[wait] (4.3,0.46) -- (4.40,0.46);
\node[lbl, anchor=north] at (4.36,0.38) {wait};
\end{tikzpicture}
\caption{Anticipatory shaping, with time running left to right. Grey bars are indexer work and blue
bars the query, and both lanes carry the same work and finish together at the dotted line. Without
shaping the query waits for the whole batch in flight; with shaping, the keystroke one debounce
interval earlier has cut the indexer work into smaller units, so the query waits for one of those.}
\label{fig:shaping}
\end{figure}

The third moment is the debounce interval, because typing precedes the query it will produce.
Each keystroke opens an activity window on the shared engine. While that window is live, the
indexer cuts its text batch into smaller units and takes the gate once per unit, and the image and
audio paths hold the gate for a single item. Less work is therefore in flight when the query arrives.
\Cref{fig:shaping} contrasts a query that waits for a full batch with one that waits for a smaller
unit. The indexer is not paused and nothing in flight is cancelled, and the longest wait a query can
inherit shrinks with the size of the unit. The smaller units carry a cost,
since taking the gate once per unit issues more kernel launches for the same work.

The store adds a second way for the indexer to slow a query down. New vectors wait in a delta outside the
base, the resident matrix the scan works from, and every query scores them separately and merges the
result. Once writes pause, the idle fold moves them into the base, and later queries no longer pay for
them. The fold, like the rest of the store maintenance, also defers for a bounded time while a search has been active recently, since it holds
the serial queue of the writer for its duration.

\subsection{The rerank funnel}
\label{sec:funnel}

A query scans the whole store: there is no approximate-nearest-neighbor index. Building one would
spend work that continuous edits keep invalidating, and its links would compete with
the encoder for memory. Search does not need exact scores for every document; it needs the right documents to reach a
shortlist, and exact scores only among those. The store therefore keeps two representations of one
space: a one-bit replica that the scan reads, and the exact vectors, which only the shortlist
touches. Both hold one row per distinct content, so a passage shared by many files is scored once.
The replica holds the sign of each coordinate, packed thirty-two to a word, so a row costs
$d/8$ bytes against $2d$ for exact bf16, a ratio of sixteen. The encoder output is a unit
vector, so a sign code needs no scale, bias or other per-row data, which makes that ratio exact.

One bit per coordinate is informative only if the coordinates carry comparable information, and two
things arrange for that. The text towers these models extend are trained with a global orthogonal
regularizer, which their authors adopted to keep retrieval quality under binary
quantization~\citep{akram2026jinav5text}. A randomized Hadamard transform is then applied to a row
before its signs are taken~\citep{zandieh2025turboquant}, since the basis an encoder happens to
produce still favors some coordinates over others. The transform is orthogonal, so it changes no
inner product and therefore no ranking; what it changes is the basis the signs are taken in,
spreading what any one dimension carries across all of them. Only the stored
side is reduced to one bit~\citep{gao2024rabitq}. The rotated query is decomposed into four bit
planes, and a Metal kernel of our own scores a row as a fixed number of population counts over its
code and those planes.

\Cref{fig:funnel} draws both representations, and marks what one query reads from each. The exact
vectors are held in a file the store maps, so the operating system keeps the rows a query gathers
and drops the rest, which costs a page fault to a local SSD when a dropped row is needed again. A
shortlist gathers rows at unrelated offsets, so the store marks the mapping \texttt{MADV\_RANDOM} to
keep the readahead of the kernel from clustering pages around every fault. What grows in memory with the index is
the replica, together with a few integer identifiers per occurrence, naming its file, its
content and its kind, and a path table that stores each directory and each file name once. The
masks a filtered query needs are derived from those identifiers on the first query that uses them.
The replica is not always present: it is adopted when the exact matrix would claim more than a
budget set from physical memory, or once the index holds more contents than a count set by the
width of the accelerator. A small index on a large machine keeps no replica and scans the exact vectors.

\begin{figure}[!htbp]
\centering
%
%
\definecolor{acc}{RGB}{31,84,140}
\begin{tikzpicture}[
  x=1.357cm, y=1cm,
  rep/.style={fill=acc!28, draw=acc!75, line width=0.5pt},
  ex/.style={fill=black!7, draw=black!45, line width=0.5pt},
  ttl/.style={font=\scriptsize, align=center},
  sub/.style={font=\tiny, text=black, align=center},
  subl/.style={font=\tiny, text=black, align=left},
  dim/.style={font=\tiny, text=black, align=left},
  ar/.style={-{Latex[length=1.3mm,width=1mm]}, draw=acc!80, line width=0.6pt},
]

\filldraw[rep] (0,0) rectangle (0.2,3.2);
\filldraw[ex]  (4.0,0) rectangle (7.2,3.2);
\filldraw[rep] (4.0,1.52) rectangle (7.2,1.68);  

\node[ttl, anchor=south west] at (-0.04,3.34) {one-bit replica};
\node[subl, anchor=north west] at (0,-0.18) {resident, read in full};
\node[dim, anchor=north west] at (0,-0.44) {$N$ rows, $d/8$ bytes each};
\node[ttl, anchor=south west] at (3.96,3.34) {exact vectors};
\node[subl, anchor=north west] at (4.0,-0.18) {file-backed, paged in on demand};
\node[dim, anchor=north west] at (4.0,-0.44) {$N$ rows, $2d$ bytes each};

\node[sub] at (5.6,2.40) {the other $N-C$ rows\\are never read};

\draw[ar] (0.4,1.6) -- (3.8,1.6);
\node[sub, anchor=south west] at (0.4,1.68) {top-$C$ by coarse score};

\draw[ar] (7.4,1.6) -- (8.4,1.6);
\node[ttl, anchor=west] at (8.6,1.66) {top $K$ files};
\node[subl, anchor=north west] at (8.6,1.52) {best exact score\\per file};
\end{tikzpicture}
\caption{The rerank funnel. Bar width is bytes per row and bar height is rows, one per distinct
content, so the replica is one sixteenth as wide and both are $N$ tall. The accent marks what one query reads: the replica in full,
the exact vectors only for the shortlist. The shortlist band is drawn thicker than $C/N$ to stay
visible.}
\label{fig:funnel}
\end{figure}

A query scans the replica, selects the $C$ best coarse scores, and rescores that shortlist against
the exact vectors, at a cost proportional to the shortlist. Each
shortlisted content is then expanded to the files that hold it, and each file keeps its best score. The coarse
scores decide only which rows enter the shortlist, so that tier does not have to be accurate,
provided the shortlist is wide enough to hold what the exact stage would have chosen. A wider code
gives a more accurate scan at any shortlist width, and one bit is the smallest replica
whose accuracy the shortlist can recover. The shortlist cannot be widened without limit, because every
additional candidate is a scattered read from the mapped file.

Once the base has supplied enough candidates, the score of the $K$-th best file bounds what any row of
the delta can contribute, and rows below that bound are skipped before their lookup. This is the
score-bound prune of document-at-a-time retrieval~\citep{broder2003}. The bound is per file, because
many rows above a row-level threshold can belong to a single file. Find similar is the same machinery
with a stored query, the mean of the vectors already stored for the picked file, so the pivot costs
one scan and no forward pass.

\subsection{Ranking across modalities}
\label{sec:rank}

The funnel returns cosine similarities, and across modalities they do not share one scale. The
encoder passes every modality through one text backbone~\citep{honicke2026jinav5omni}, and against a
text query an unrelated photograph scores like an unrelated document; a relevant photograph, however,
scores below a relevant document. \app{} keeps one ratio $s$ per encoder for images, audio and video
against text: the median score a caption gives its own image over the median it gives the other
captions of that image. A text query ranks every media chunk by its score divided by $s$. The division is applied where the
scan reduces chunks to files, and every candidate selection before that, the coarse selection of the
funnel included, ranks on the divided score. One scan therefore returns the exact top $k$ of the scaled
scores, and no media answer is cut before it is scaled. Each chunk is scaled by its own modality, so a
screenshot competes by its picture and by its text. A query that is not text, such as a search for
files similar to a given one, is left unscaled. The rule uses no word list and decides nothing per
query, so it applies in every language the encoder reads. The same scale decides what is shown: the
relevance threshold the user sets applies to the scaled score, and a hit whose scaled score is below
0.6 of the best in the list is folded under a single control, apart from the first hit.

Two properties of a file are absent from its vectors: its name, and the identifiers it contains,
such as an invoice number, a booking code or a date. Filename search uses a small full-text index
over basenames, fused with the dense results only when the query looks like a name, carrying an
extension or consisting of a few uncommon tokens. The fused score adds a weighted name-match
strength in $[0,1]$ to the cosine similarity, so one relevance threshold holds for fused and unfused
queries alike, and a query that is the whole name of a file ranks that file first. An identifier is a run of four to forty ASCII letters, digits and joining marks that
holds a digit and either a letter or seven digits in all. The indexer collects identifiers from each
chunk as it cuts the text and keeps them in a full-text sidecar, one document per chunk, keyed by
the content digest of the file. A rename or a copy therefore needs no write to the sidecar, and a deleted file leaves a key
that resolves to no file and is swept later. A query made only of identifiers puts the files that
hold all of them first, in the order of their dense scores, and opens each at the chunk holding the
first; a query with words in it is left to the vectors.

\subsection{Durability and deletion}
\label{sec:store}

The index is derived state, and durability follows from that. The database is kept in
write-ahead-logging mode at normal synchronization, so a crash can cost the last transactions, and
the cost is a rescan of the affected files. The vectors live in the mapped file, and the database
separates what a chunk is from where it occurs. A content table holds each distinct chunk once,
keyed by its hash, with its kind, a count of its occurrences and the position of its vector in the
file. An occurrence table holds one row per chunk of each file, pointing at a content and carrying
the locator. Snippets sit in a table keyed by content, which a cold load never reads and a query
reads only for the results it returns. Identity and position are separate columns, so a vector
can move in the file while its content keeps its identity. A monotone counter over mutations is
written inside the transaction of the mutation it describes. The cached copy of the resident row
table carries that counter as a stamp, so a copy that predates a committed change is rejected at open and rebuilt
from the tables without re-encoding.

Deletions and renames arrive as ordinary watcher events: a path that no longer resolves has its rows
removed. A move is handled the same way for its old half, after the new path has taken the
vectors of the old rows, so a renamed file is not re-encoded.
Removing a file moves no other row. Its occurrences are marked dead in memory and skipped by every
scan, and each content it pointed at loses a reference. A content left with none releases its
position, which is handed to the next new content once the releasing transaction commits, because a
query in flight may still hold it. The exact file and the one-bit replica are not written in the
same instant, so the transaction that reuses a position also records it. After an abrupt exit, the
next open reconciles both representations at the recorded positions alone, and no row is scored with
the vector of the content that held its position before.

Queries and writes reach the store through one serial queue, so the longest hold of that queue
bounds how long a query can wait. Some changes touch a large part of the index at once: a folder
removed from the search scope, a file type turned off, or many files rewritten after a change of
settings. Each is applied in slices, every slice a complete transaction sized from the measured cost
per row to hold the queue for about a tenth of a second, so a query that arrives during the change
waits for one slice at most. On a cold page cache most of a slice is spent reading pages, so those
pages are read beforehand on separate read-only connections, outside the queue. Free positions are
reclaimed in the same way. The vectors at the end of the file move into free positions at its start,
a few at a time, each move committed together with the position it frees. Once every free
position lies past every position in use, the file is cut there. No step rewrites the whole vector
file, and an interruption at any point leaves a consistent index. Deleting rows leaves the database
file at its high-water mark, so free space is reclaimed by a compaction that runs only when enough
of the file is free to justify rewriting it.

\subsection{Memory budget}
\label{sec:cap}

\app{} stays open and keeps indexing while the user works, so it holds memory for as long as the
machine is in use. It therefore exposes a single memory setting. What must stay resident, the model
weights and the scan matrix, is not for the user to limit. A total set below it cannot be honored,
and the array framework then contends for its allocator on every allocation while the process
occupies the same memory anyway. The setting is therefore headroom, the memory the process may hold
beyond that resident set, which goes to larger batches and a buffer cache.
macOS provides no process-wide limit either: it reports an approximation of what a device can
allocate before its performance suffers, and that approximation is
advisory~\citep{apple2026devicememory}. Keeping the headroom takes more than bounding the allocator
the setting names, because its neighbors draw on the same memory and the system swaps against their
sum.

That one user-visible number is therefore the input to each allocator found to draw on unified
memory. The array framework is limited to the resident set plus a working floor and the headroom.
A byte gate bounds how much decoded media may be resident at once. The packing budget for vision
inputs derives from the budget, as does the page cache of the embedded database. So does the buffer
cache of the array framework, which is the easiest of these to miss because no line of application
code allocates it. No interface enumerates the allocators that draw on the pool, so the list has to
be assembled by measurement.

One number per allocator is also not always sufficient. The page cache of the embedded database is sized for bulk insert, but a compaction
rewrites the whole file through it with the model weights and the base resident. The cache is
therefore shrunk for the duration of a compaction and restored afterwards.

The same reasoning applies to transients. Converting the scan matrix to a different representation
will, done naively, materialize a full-height destination beside the source. That is precisely the
doubling the one-bit replica was adopted to avoid, and the buffer cache then retains the
transient after the conversion returns. We perform the conversion in slabs instead. The same cache
makes slabbing cheap, since each slab reuses the buffer of the one before it. Decoded video frames
are staged as bytes for the same reason, at a quarter of the size of 32-bit floats, because the
budget must hold at the peak reached during decoding.

\subsection{The Neural Engine}
\label{sec:ane}

Apple silicon carries a Neural Engine beside the GPU, an accelerator that runs compiled Core ML
programs in fp16~\citep{orhon2022ane}. When the user chooses it, \app{} moves all of its work off
the GPU: every forward pass runs on the Neural Engine and every scan of the store on the CPU, and no
work is split between the two accelerators. The GPU is then left to the other applications, and the
indexer no longer competes with them for it.

A Neural Engine program is compiled for fixed shapes, by one compiler that serves the whole machine
at a cost that grows with the size of the graph. The text backbone is one program per model, with a
function per sequence length and one copy of the weights shared by all of them; that copy is written
on the device from the weights the application already holds, so nothing further ships. The image
and audio towers are programs of one layer each that take the weights as inputs, so each compilation
covers a single layer at one input size. Every function is compiled once, when the user
turns the mode on, and later launches load the compiled form. A sequence longer than the longest
text function, as a video segment produces, runs one layer per call over the weight file of the text
program. The GPU copies of the towers are released, so no weight is held twice.

The arithmetic is fp16 throughout, and a reduction over a large activation can overflow it. In
\modelsmall{}, the attention sink~\citep{xiao2024sinks}, a token whose activations far exceed those
of the others, makes the sum of squares in the root-mean-square norm overflow fp16, so the program
divides each row by its largest magnitude before the norm, which leaves the result unchanged. Short
sequences, such as the labels of the tag vocabulary, share one call under a block-diagonal mask, with
positions restarting for each. A query waiting on the Neural Engine is let in before the indexer
takes its next call, as at the gate.

In this mode the store scans on the CPU. The one-bit scan becomes SIMD population counts over the
replica, which is integer arithmetic, so the CPU reproduces the scores of the GPU kernel to the bit.
The shortlist is selected exactly, and its rows are rescored by a SIMD dot product over the bf16
vectors. On the CPU the replica overtakes the exact scan at a smaller
index than on the GPU, so this mode adopts it at a lower row count.

\section{Evaluation}
\label{sec:eval}

We measure cost on one workload that \app{} generates from a seed, so each machine runs the same
work on the same bytes. It holds text files, images, and audio and video clips for the tasks
that run the encoder, and a store of 946{,}568 chunks of seeded vectors for the tasks that search or
change an index. A task that also indexes writes the generated files into a copy of that store.
The store is laid out as a personal index is: most files are short and a few are long, one file in
eight is an image, a tenth of all passages repeat across files, and the files sit in a tree of
folders. Each task that searches or changes the store runs on its own copy, opened and read
through once as the application reads its index at launch. We run a task on every machine whose
physical memory holds twice its measured peak, and show a dash otherwise.

\subsection{Machines}
\label{sec:eval-setup}

\begin{table}[!htbp]
\centering
\caption{The five machines and the representation the one-million-chunk store adopts on each,
which follows from the width of the accelerator.}
\label{tab:machines}
\footnotesize
\setlength{\tabcolsep}{4pt}
\begin{tabular}{lrrrrr}
\toprule
Quantity & \shortstack{M3 Ultra\\2025} & \shortstack{M4 Pro\\2024} & \shortstack{M4\\2024}
         & \shortstack{M3 Pro\\2023} & \shortstack{M2\\2022} \\
\midrule
CPU cores (performance + efficiency) & 24 + 8 & 10 + 4 & 4 + 6 & 5 + 6 & 4 + 4 \\
GPU cores                            & 80     & 20     & 10    & 14    & 10 \\
Unified memory (GB)                  & 512    & 48     & 16    & 18    & 16 \\
macOS                                & 26.6.2 & 26.7.1 & 15.7.3 & 26.5.0 & 26.6.0 \\
Scan representation                  & exact & one bit & one bit & one bit & one bit   \\
\bottomrule
\end{tabular}
\end{table}

\Cref{tab:machines} describes the five machines. Every latency, throughput and memory measurement
below comes from \app{} with \modelnano{} at the 1\,GB of headroom it ships with, except the
comparison of the two accelerators in \Cref{sec:eval-ane}, which runs both encoders. We measure on
mains power with the applications of each machine left running, as \app{} runs in daily use. For
each task we record the share of the CPU and the GPU taken by other processes, the slowest clock,
the swap read back and the thermal state, which stays nominal throughout. We compare each mechanism
with the same binary with that mechanism disabled. The papers of the encoder report its retrieval quality on
standard benchmarks~\citep{akram2026jinav5text,honicke2026jinav5omni}, so the question here is what
each mechanism costs and saves on a device that also runs everything else, and what the ranking
built on the encoder retrieves from local files.

\subsection{Task latency}
\label{sec:eval-tasks}

Every row of \Cref{tab:tasks} is one task measured end to end through the path the application
itself takes. A query encodes what the user supplied and then searches, and the two halves depend
on different things: encoding time on the input, search time on the store and the device.
Encoding a short query costs between 2.8 and 6.4\,ms, since one query is one small forward pass,
and the search over the same store between 4.1 and 10.9\,ms. Encoding recurs on every keystroke,
which is why the query path debounces it, and the search is the half that grows with the index, which is why the funnel
shrinks it.

Find similar starts from a vector already stored and skips encoding, so it costs 33 to 43\% less
than a text query. A query made of an identifier adds the lookup in the sidecar to a text query,
0.5 to 0.8\,ms, and puts the file that holds it first for all 120 such queries. A query whose
answers are pictures searches in the time any other query does, since the media scale is applied
within the same scan. A filtered query costs what an unfiltered search costs
once the mask columns exist, and the first query after a filter changes, which builds them, takes
4.3 to 12.0\,ms. The media queries are the opposite case, dominated by decoding and by a tower that runs once
per item, and cost between 30 and 358\,ms at the median. Open-vocabulary tagging changes the cost of indexing an image by 3.3\% at most, since the tags come from a matrix
multiplication against the resident label matrix on a pass the image already required.

The search-while-indexing row measures shaping in the condition it is built for. An indexing load
writes into the store being searched, with chunk reuse off so that every file it indexes is
embedded again. A query therefore waits both for the accelerator and for the new vectors the scan
has to cover, and each query follows a keystroke and the debounce interval, so the
mechanism is armed exactly as it is in use. Shaping bounds the largest unit a query can arrive
behind, so what it moves is the worst case: on the machines with 10 to 20 GPU cores the 99th
percentile falls by 49 to 58\%, while indexing throughput over the same window stays within 2.0\% of
the unshaped run. On the M3 Ultra, whose batches already finish
quickly, shaping takes 22\% off the 99th percentile for 4.6\% of indexing throughput.

The last block measures a search issued every 50\,ms while one bulk change runs on a copy of the
store: 5{,}000 files rewritten, 50{,}000 files deleted, one of eight top-level folders removed, every
image removed, and the space those removals free reclaimed. Each change holds the queue one slice at
a time, so the longest wait stays under half a second on every machine while the changes themselves
take up to 7.2\,s, and \Cref{sec:eval-bulk} lists their durations.

\begin{table}[!htbp]
\centering
\caption{What each task costs in milliseconds, as p50 / p95 / p99 nearest-rank percentiles. Query
rows and the last block search the one-million-chunk store, and the filename and filtered rows time
the search alone. The indented row is the first query after a filter changes, which builds the
mask columns. A cell without a p99 has under 100 samples.}
\label{tab:tasks}
\scriptsize
\setlength{\tabcolsep}{3.5pt}
\begin{tabular}{lrrrrr}
\toprule
Task & \shortstack{M3 Ultra\\2025} & \shortstack{M4 Pro\\2024} & \shortstack{M4\\2024}
     & \shortstack{M3 Pro\\2023} & \shortstack{M2\\2022} \\
\midrule
Filename query   & 4.6 / 5.8 / 6.2 & 6.5 / 6.8 / 7.0 & 8.8 / 9.1 / 9.9 & 7.9 / 8.7 / 9.2 & 11.7 / 13.7 / 14.3 \\
Text query       & 7.0 / 8.2 / 8.7 & 9.2 / 9.6 / 10.6 & 13.3 / 13.7 / 14.5 & 12.1 / 12.4 / 13.6 & 17.4 / 18.6 / 19.1 \\
Identifier query & 7.7 / 8.9 / 9.4 & 9.9 / 10.4 / 10.6 & 13.8 / 14.6 / 15.2 & 12.6 / 13.3 / 13.7 & 18.2 / 20.6 / 23.0 \\
Filtered query   & 4.1 / 4.6 / 5.3 & 5.6 / 5.9 / 6.0 & 8.5 / 8.7 / 9.5 & 7.4 / 7.6 / 7.8 & 11.1 / 12.2 / 13.3 \\
\quad first of a filter & 4.3 & 7.9 & 10.7 & 9.5 & 12.0 \\
Find similar     & 4.0 / 4.7 / 5.4 & 6.2 / 6.4 / 6.8 & 8.4 / 9.0 / 9.3 & 7.5 / 7.6 / 8.1 & 10.8 / 11.8 / 12.8 \\
Image query      & 30.5 / 32.2 & 67.8 / 71.1 & 128 / 130 & 105 / 113 & 187 / 213 \\
Audio query      & 73.1 / 91.9 & 180 / 244 & 358 / 491 & 279 / 401 & --- \\
Video query      & 98.4 / 104 & 94.5 / 103 & 186 / 193 & 163 / 171 & 229 / 395 \\
Index one image  & 15.9 / 31.6 & 33.3 / 67.9 & 62.3 / 125 & 50.2 / 99.7 & 83.9 / 177 \\
Save one edit    & 9.0 / 13.1 / 15.1 & 15.3 / 25.1 / 31.9 & 21.2 / 36.5 / 42.8 & 18.7 / 33.3 / 38.9 & 31.4 / 56.0 / 70.8 \\
Search while indexing & 70.3 / 97.7 / 130 & 71.5 / 118 / 132 & 115 / 212 / 224 & 95.6 / 181 / 195 & 193 / 316 / 349 \\
\midrule
Search, no write running & 4.4 / 5.8 / 8.0 & 13.0 / 15.2 / 16.6 & 9.0 / 11.0 / 12.0 & 14.1 / 16.1 / 17.1 & 16.4 / 21.3 / 24.7 \\
\quad rewriting files    & 4.8 / 35.8 & 13.4 / 21.9 & 10.2 / 19.9 & 18.0 / 35.6 & 18.7 / 36.3 \\
\quad deleting files     & 4.4 / 110 / 111 & 13.4 / 102 & 13.6 / 121 & 15.1 / 115 & 17.0 / 114 / 140 \\
\quad removing a folder  & 4.2 / 89.6 & 15.0 / 73.1 & 12.4 / 80.6 & 15.5 / 77.3 & 19.7 / 113 \\
\quad removing a file type & 4.6 / 91.4 & 14.1 / 61.5 & 11.5 / 56.4 & 17.2 / 79.3 & 19.9 / 63.3 \\
\quad reclaiming space   & 4.7 / 72.0 & 13.5 / 55.0 & 10.5 / 58.0 & 17.3 / 60.8 & 17.7 / 75.0 \\
\bottomrule
\end{tabular}
\end{table}

\subsection{Mechanism ablations}
\label{sec:eval-main}

\begin{table}[!htbp]
\centering
\caption{Indexing throughput and what each mechanism saves.}
\label{tab:main}
\footnotesize
\setlength{\tabcolsep}{4pt}
\begin{tabular}{lrrrrr}
\toprule
Quantity & \shortstack{M3 Ultra\\2025} & \shortstack{M4 Pro\\2024} & \shortstack{M4\\2024}
         & \shortstack{M3 Pro\\2023} & \shortstack{M2\\2022} \\
\midrule
Index throughput (tokens/s)        & 78{,}452 & 25{,}769 & 13{,}027 & 16{,}069 & 8{,}694 \\
\quad per GPU core                 & 981 & 1{,}288 & 1{,}303 & 1{,}148 & 869 \\
Bulk write (chunks/s)              & 26{,}897 & 32{,}409 & 32{,}309 & 28{,}764 & 24{,}828 \\
\midrule
Score-bound prune (\% latency saved) & 22.4 & 18.8 & 12.5 & 16.3 & 17.3 \\
Idle fold (\% latency saved)       & 37.0 & 16.9 & 22.6 & 24.9 & 10.9 \\
Shaping (\% search p99 saved)      & 21.6 & 48.5 & 58.0 & 49.7 & 55.6 \\
Shaping (\% indexing throughput lost) & 4.6 & $-$0.3 & $-$0.8 & $-$2.0 & $-$0.7 \\
\bottomrule
\end{tabular}
\end{table}

\Cref{tab:main} reports each mechanism, with the enabled and disabled arms interleaved run by run
against the same store. Indexing time is spent in the encoder, and more so on a
narrower device: the accelerator is busy for 95 to 99\% of a fresh pass. Throughput per GPU
core spans 869 to 1{,}303 tokens per second across an eightfold range of accelerator width, so the effort belongs in the forward passes themselves and in which path reaches
the accelerator first. Tail-row narrowing gains 5.1\% of throughput, close to the 6.25\% the
architecture implies, and narrow operands run fused attention 1.19 to 1.31$\times$ faster than fp32
at the shapes batched images produce, as \Cref{sec:eval-arith} details. Reusing the unchanged chunks
of an edited file cuts the median cost of saving it by 75 to 88\%. Ordering the first pass by
turns makes every file of a tree with long files searchable after 48\% of the pass on every machine,
between 3.5 and 31.4\,s into a pass of 7.3 to 65.8\,s.

On the query path the idle fold saves 11 to 37\% of a query and the score-bound prune 13 to 22\%.
\Cref{sec:eval-scale} gives the recall the funnel keeps, 95.8\% of an exact top 10 over 500{,}000 rows
and 88.3\% over eight million, and \Cref{sec:eval-select} gives a selection of its shortlist that runs up to
9.5$\times$ faster than the primitive the framework offers. \Cref{tab:cross} shows that the width of
the accelerator decides at what size the replica becomes the faster scan. The machines with 10 and 14
GPU cores cross over by 500{,}000 rows, the 20-core M4 Pro by one million and the 80-core M3 Ultra
between one and two million. Marking the rows of a removed file dead takes 0.32\,ms at 500{,}000 rows
on the M3 Ultra, against 453\,ms for compacting the resident matrix, as \Cref{sec:eval-delete} details.

Bounding the page cache for the compaction of a 186\,MB index holds its peak at 176\,MB above the
footprint before it starts, against 200\,MB unbounded. Bounding each transient does not by itself keep
their sum inside the budget, so we also trace the whole process while every kind of file is indexed
into the store as it is searched, and while a bulk deletion and a compaction follow. Its peak stays
within 2.9\% of what the setting allows, and below it on the machines with 16 and 18\,GB, as
\Cref{tab:memory} details.

\subsection{Retrieval quality}
\label{sec:eval-quality}

We measure retrieval on a labeled testbed generated from a fixed specification and from the
pictures and offline voices built into macOS. Its 226 files include documents in ten languages,
some beside a near twin that differs in one detail, PDFs with a different topic on each page, a
scanned page, voice memos, a ten-minute lecture, videos, photographs, screenshots and two long logs
that each hide one fact, among 120 distractor notes. Each of 125 queries names the file that answers
it and, where the answer has one, its place: a page, a line range or a moment. A hit counts only at
the right place, and we report the mean reciprocal rank (MRR) of the first such hit in the list the
application returns. The testbed is indexed and queried through the indexer and search path of
\app{}, with both sizes of the encoder.

\begin{table}[!htbp]
\centering
\caption{Mean reciprocal rank on the labeled testbed with one step removed at a time. The group is
the queries the step serves, given as shipped / removed; all is the 125 queries with the step
removed, against 0.896 and 0.925 shipped. The last row removes these steps at once, together with
screenshot text and filename fusion.}
\label{tab:quality}
\scriptsize
\setlength{\tabcolsep}{3.5pt}
\begin{tabular}{llr>{\centering\arraybackslash}p{2.45cm}>{\raggedleft\arraybackslash}p{1.0cm}%
                    >{\centering\arraybackslash}p{2.45cm}>{\raggedleft\arraybackslash}p{1.0cm}}
\toprule
& & & \multicolumn{2}{c}{\modelnano{}} & \multicolumn{2}{c}{\modelsmall{}} \\
\cmidrule(lr){4-5}\cmidrule(lr){6-7}
Removed step & Queries it serves & $n$ & group & all & group & all \\
\midrule
Identifier lookup & an identifier alone & 15 & 1.000 / 0.552 & 0.843 & 1.000 / 0.603 & 0.877 \\
Modality scale & pictures and recordings & 32 & 0.732 / 0.556 & 0.852 & 0.811 / 0.700 & 0.901 \\
Soundtrack & speech in a video & 3 & 1.000 / 0.000 & 0.873 & 1.000 / 0.014 & 0.901 \\
Page cut of short PDFs & a page of a short PDF & 4 & 1.000 / 0.436 & 0.882 & 1.000 / 0.525 & 0.913 \\
\midrule
All of the above & all & 125 & & 0.766 & & 0.812 \\
\bottomrule
\end{tabular}
\end{table}

\Cref{tab:quality} removes one step at a time from the shipped search and scores both the queries
the step serves and the whole set. Every removal lowers the score of the queries it serves.
The identifier lookup matters most, because the encoder scores a document that names a code much like
any other document. Without the page cut, a question about one page of a short PDF
still finds the document but ranks it lower. Without the soundtrack, the moment a query about speech asks for falls out of the first twenty
results. Removing every step, which leaves cosine similarity
over the chunks alone, lowers MRR over all queries from 0.896 to 0.766 with \modelnano{} and from
0.925 to 0.812 with \modelsmall{}. The weaker-match fold leaves 4.3 and 4.5 of the 120 hits a query
returns in view, and over both encoders it folds one answer that ranks within the first ten.

A second testbed adds three presentations, six spreadsheets and three PDFs that each hold a page
that is mostly a picture, with 30 queries about them. Six of these queries are answered only by the
picture of a slide or of such a page. With the pictures indexed, all six rank their answer first
with both encoders; without them, the three about slides find nothing and the three about PDF pages
score an MRR of 0.029 and 0.024.

The ratio $s$ comes from 2{,}000 images of COCO~\citep{lin2014coco} and their 10{,}000 captions,
embedded through the same paths. Against a caption, an unrelated image scores like an unrelated
caption, 0.120 against 0.107 with \modelnano{} and 0.127 against 0.095 with \modelsmall{}. The
image a caption describes scores 0.427 and 0.480 against 0.648 and 0.626 for its other captions,
which gives $s = 0.66$ and $s = 0.77$. In a pool of the images and the other captions, the image a
caption describes ranks 186th and 82nd at the median unscaled, and 13th and 9th scaled.

\subsection{The Neural Engine}
\label{sec:eval-ane}

\Cref{tab:ane} compares the GPU and the Neural Engine on the M3 Ultra with both encoders. The two runs follow each other
with no other work on the GPU, so the only GPU load beside \app{} is an application of our own, a
fixed Metal workload at 60 frames per second sized to a quarter of each frame. Power is the draw of
the whole system while text is indexed, less its draw at idle.

\begin{table}[!htbp]
\centering
\caption{Both encoders on the M3 Ultra, on the GPU and with the encoder on the Neural Engine (NE)
and every scan on the CPU. Latencies are p50 / p95 in milliseconds, searches run over the
one-million-chunk store, and own GPU time leaves out the tasks that load the GPU by design.}
\label{tab:ane}
\footnotesize
\setlength{\tabcolsep}{3.5pt}
\begin{tabular}{l*{4}{>{\raggedleft\arraybackslash}p{2.0cm}}}
\toprule
& \multicolumn{2}{c}{\modelnano{}} & \multicolumn{2}{c}{\modelsmall{}} \\
\cmidrule(lr){2-3}\cmidrule(lr){4-5}
Quantity & GPU & NE & GPU & NE \\
\midrule
Index throughput (tokens/s)          & 78{,}279 & 39{,}623 & 19{,}491 & 9{,}303 \\
Index one image (ms)                 & 13.0 / 26.0 & 38.0 / 79.9 & 36.2 / 69.5 & 117 / 239 \\
First pass over a tree of files (s)  & 7.3 & 14.6 & 29.1 & 62.7 \\
Image query (ms)                     & 27.1 / 28.0 & 77.3 / 83.9 & 71.8 / 73.0 & 236 / 243 \\
\midrule
Search, no write running (ms)        & 4.7 / 7.1 & 4.0 / 13.4 & 4.9 / 11.6 & 4.2 / 14.1 \\
Search while indexing (ms)           & 65.9 / 94.8 & 5.8 / 7.2 & 43.3 / 142 & 5.7 / 7.7 \\
Search beside the GPU application (ms) & 7.7 / 12.3 & 7.4 / 17.4 & 11.6 / 25.3 & 7.2 / 17.9 \\
Index throughput lost beside it (\%) & 18.2 & $-$5.7 & 19.8 & $-$0.2 \\
Own GPU time over the run (s)        & 143 & 2.3 & 311 & 2.3 \\
\midrule
Power while indexing text (W)        & 134 & 20.7 & 139 & 20.2 \\
Energy per million tokens (kJ)       & 1.71 & 0.52 & 7.15 & 2.18 \\
Memory at rest (MB)                  & 6{,}108 & 5{,}328 & 7{,}194 & 7{,}261 \\
Peak memory, indexing every kind (MB) & 8{,}412 & 5{,}510 & 9{,}712 & 7{,}601 \\
\midrule
MRR on the labeled testbed           & 0.896 & 0.905 & 0.925 & 0.922 \\
\bottomrule
\end{tabular}
\end{table}

The Neural Engine indexes text at half the throughput of the GPU with both
encoders, and indexing an image or searching with one takes about three times as long. The scan keeps its
speed on the CPU, where a search with nothing else running costs 4.0 and 4.2\,ms at the median, and
the replica overtakes the exact scan there at a smaller index, as \Cref{sec:eval-cross} details. In return the
mode leaves the GPU free. Over a run the process itself uses 2.3\,s of GPU time with either encoder, against 143 and
311\,s with the encoder on the GPU. The GPU application then takes none of the indexing throughput,
where it takes 18 and 20\% on the GPU. A search issued while the indexer runs costs under 6\,ms
at the median against 66 and 43\,ms, since a scan on the CPU waits for no accelerator work. The
system draws about 20\,W while it indexes text against 134 and 139\,W, so each token costs under a
third of the energy although it is encoded more slowly. Peak memory falls by 2.1 to 2.9\,GB. The
index built in this mode scores an MRR of 0.905 and 0.922 on the labeled testbed, against 0.896 and
0.925 when built on the GPU.

\section{Related work}
\label{sec:related}

Systems that serve foundation models under memory pressure have concentrated on the server case.
vLLM \citep{kwon2023} pages the KV cache to raise batch occupancy, while on a single-user desktop
the scarce resident structure is the vector store. Clockwork
\citep{gujarati2020} argues for predictable inference by eliminating choice inside the serving
stack, which is the principle behind our gate; on a device with one accelerator and no replicas that
principle reduces to admission control. Flash-resident inference \citep{alizadeh2023} shares our
premise that memory is the binding constraint on a personal device, but streams the weights of a
model that does not fit, while our model fits and the pressure comes from the index and from
concurrent consumers. A comparison of local inference runtimes on large-memory Apple
silicon \citep{rajesh2025prodlocal} measures the serving path on its own. Apple describes a layout
that runs a transformer efficiently on the Neural Engine~\citep{orhon2022ane}, which our text
backbone follows.

On the retrieval side, product quantization \citep{jegou2011} established the coarse-then-exact
structure used here, and graph indexes \citep{malkov2018} together with GPU implementations of
quantization and graph search \citep{johnson2017} are the standard way to avoid scanning every
vector. \app{} uses neither. Storing embeddings at low precision for retrieval is established
\citep{jeong2025int4rag}, and locally-adaptive quantization reaches its best rates by fitting scales
per vector, which has to be re-established once the collection
streams~\citep{aguerrebere2024lvq,adenali2025quantstream}. A rotation applied before quantization
removes the outlier coordinates that cost an affine code its precision~\citep{zandieh2025turboquant}.
A rotated sign code scored asymmetrically, with the query held at a higher precision than the
stored side, is accurate enough to carry the coarse tier of a funnel~\citep{gao2024rabitq}. The
replica here is that construction, and unlike an affine code it holds no per-row parameters at all,
so the code of a row depends on that row alone and adding a row codes nothing else. The
multi-vector case is compressed by clustering and pruning patch embeddings before late
interaction~\citep{bach2025hpccolpali}, a pressure that does not arise where one chunk contributes
one vector.

Omni-modal encoders that embed several modalities into one representation space
\citep{girdhar2023,zhu2023,xu2025omniembed,tonmoy2026fusion} are the enabling model work, and we
consume such an encoder. Contrastive training leaves each modality in a cone of its own, which is
known as the modality gap~\citep{liang2022mindgap}. The encoder used here shows no such offset between
unrelated items yet scores relevant media below relevant text, and \app{} corrects that difference in
score scale at query time and leaves the encoder unchanged. Local-first retrieval has been argued as a design position and benchmarked
on consumer hardware \citep{zerhoudi2026aswemaysearch}, for its retrieval quality and without a
process that indexes and answers queries at the same time.

Two on-device systems are closest. Storage-efficient vector indexing \citep{wang2025leann} shares
the deployment target and optimizes index size: it recomputes embeddings and
stores none of them, and reports far smaller indexes as a result. We keep the vectors because the
encoder is already resident and saturated during indexing, so an index that recomputes would put
encoder forward passes on the query path. Content-hash reuse of encoder work appears in native
multimodal serving on this hardware \citep{barrios2026nativeapple}, where a whole-object cache is
the right unit for repeated inputs. Our unit is a content-defined chunk, a technique introduced to
deduplicate network file-system traffic~\citep{muthitacharoen2001,xia2016}, keyed across the whole
index, so the unchanged part of an edited file and a passage repeated across files cost no forward
pass. A whole-object cache misses both.

\section{Conclusion}
\label{sec:conclusion}

To our knowledge, \app{} is the first engine to run an omni-modal encoder in the application process
on consumer Apple silicon while a continuous indexer and an interactive query path share the device.
Once search moves onto the machine that holds the files, the encoder stops being the hard part.
Indexing never finishes, so what has to stay cheap is the next edit, and addressing chunks by their
content makes its cost follow the text that changed. A first pass still takes hours, and a small
first turn for each file makes every file searchable early in that pass. One representation space
still scores relevant media below relevant text, and one ratio per encoder, applied inside the scan,
puts every modality in one list. The device is shared, yet committed GPU work
cannot be preempted, so a query can be favored only where no indexer work is in flight and in the
moment before it is issued. A second accelerator offers another way to share it: with the encoder on
the Neural Engine and the scan on the CPU, the indexer leaves the GPU to the applications beside it
and spends under a third of the energy per token.
Unified memory makes every allocator draw on the headroom the user sets, including allocators no setting names. These constraints bind hardest on the weakest devices, which is where
admission control and the one-bit replica save the most. Runtimes that expose preemption and one
process-wide memory budget to applications would let any on-device AI application share the device
this way without rebuilding these mechanisms.

\bibliographystyle{plainnat}
\bibliography{refs}

\clearpage
\appendix
\crefalias{section}{appendix}
\crefalias{subsection}{appendix}
\setcounter{figure}{0}\renewcommand{\thefigure}{A\arabic{figure}}\renewcommand{\theHfigure}{A\arabic{figure}}
\setcounter{table}{0}\renewcommand{\thetable}{A\arabic{table}}\renewcommand{\theHtable}{A\arabic{table}}
\section{Appendix}
\label{sec:appendix}

\subsection{Bulk changes}
\label{sec:eval-bulk}

\Cref{tab:bulk} gives what each bulk change takes and what a search beside it waits.

\begin{table}[!htbp]
\centering
\caption{Each bulk change on a copy of the one-million-chunk store, given as its duration in seconds
/ the longest wait in milliseconds of a search issued every 50\,ms while it runs. The folder is one of
eight top-level folders, and reclaiming follows the removal of every image.}
\label{tab:bulk}
\footnotesize
\setlength{\tabcolsep}{4pt}
\begin{tabular}{lrrrrr}
\toprule
Change & \shortstack{M3 Ultra\\2025} & \shortstack{M4 Pro\\2024} & \shortstack{M4\\2024}
       & \shortstack{M3 Pro\\2023} & \shortstack{M2\\2022} \\
\midrule
Rewrite 5{,}000 files & 2.3 / 96 & 1.5 / 217 & 1.5 / 272 & 1.8 / 117 & 2.7 / 453 \\
Delete 50{,}000 files & 5.2 / 115 & 4.2 / 112 & 4.8 / 153 & 4.6 / 129 & 7.2 / 232 \\
Remove a folder & 1.6 / 117 & 1.2 / 97 & 1.5 / 100 & 1.5 / 120 & 2.2 / 160 \\
Remove every image & 1.2 / 96 & 1.0 / 109 & 1.2 / 112 & 1.2 / 111 & 2.0 / 137 \\
Reclaim the freed space & 2.4 / 99 & 1.6 / 139 & 1.7 / 104 & 2.0 / 82 & 4.3 / 112 \\
\bottomrule
\end{tabular}
\end{table}

\FloatBarrier
\subsection{Attention operand precision}
\label{sec:eval-arith}

\begin{figure}[!htbp]
\centering
%
%
\begin{tikzpicture}
\pgfplotsset{every axis/.append style={font=\scriptsize, ymajorgrids,
  major grid style={draw=black!12}, axis lines*=left, tick align=outside, tick pos=left}}

\begin{axis}[
  width=\linewidth, height=4.6cm,
  xlabel={patch sequence length}, ylabel={fused attention (TFLOPS)},
  xmode=log, log basis x=2, xmin=200, xmax=6000,
  ymode=log, ymin=0.2, ymax=80,
  ytick={0.25,1,4,16}, yticklabels={0.25,1,4,16},
  xtick={256,512,1024,2048,4096}, xticklabels={256,512,1k,2k,4k},
  legend style={font=\tiny, draw=none, fill=none, at={(0.01,0.99)}, anchor=north west},
  legend columns=3, legend style={/tikz/every even column/.append style={column sep=6pt}},
  legend cell align=left,
  every node near coord/.append style={font=\tiny},
]
  \addplot[mark=*, mark size=1.5pt, thick, black!75] coordinates {
    (256,1.05) (512,3.49) (1000,8.75) (1272,10.58) (2000,15.26) (4888,20.19)};
  \addlegendentry{M3 Ultra, 80 GPU cores}
  \addplot[mark=square*, mark size=1.4pt, thick, black!60, densely dotted] coordinates {
    (256,1.35) (512,3.21) (1000,4.30) (1272,5.32) (2000,5.85) (4888,6.47)};
  \addlegendentry{M4 Pro, 20 GPU cores}
  \addplot[mark=triangle*, mark size=1.8pt, thick, black!50, dashed] coordinates {
    (256,1.15) (512,2.42) (1000,2.93) (1272,3.30) (2000,3.64) (4888,3.99)};
  \addlegendentry{M3 Pro, 14 GPU cores}
  \addplot[mark=diamond*, mark size=1.6pt, thick, black!40, dash pattern=on 3pt off 1.5pt] coordinates {
    (256,0.69) (512,1.72) (1000,2.37) (1272,2.72) (2000,3.00) (4888,3.28)};
  \addlegendentry{M4, 10 GPU cores}
  \addplot[mark=pentagon*, mark size=1.5pt, thick, black!35, dash pattern=on 1pt off 1.5pt] coordinates {
    (256,0.72) (512,1.27) (1000,1.72) (1272,1.94) (2000,2.07) (4888,2.21)};
  \addlegendentry{M2, 10 GPU cores}
  \draw[dashed, black, line width=0.6pt] (axis cs:1000,0.2) -- (axis cs:1000,24);
  \node[font=\tiny, anchor=north west, align=left] at (axis cs:1080,0.62)
    {one image\\$\approx$1000 patches};
\end{axis}

\end{tikzpicture}
\caption{Fused attention throughput with bf16 operands against sequence length at the attention
shape of the vision tower, on a logarithmic scale. The dashed line marks one image, which is one forward pass.}
\label{fig:latency}
\end{figure}
\Cref{fig:latency} places one image, at about a thousand patches, at 43\% of the throughput of the
longest shape on the widest machine and at up to 78\% on a narrow one. At the long shapes that batched
images produce, 2{,}000 and 4{,}888 patches, bf16 operands with fp32 accumulation are 1.19 to
1.31$\times$ faster than fp32 operands on every machine, and the tower spends its time at those
shapes. The fp16 operands the tower uses run at 0.94 to 1.03 times the throughput of bf16 at the
six shapes on the M3 Ultra, so their three more bits of mantissa cost little speed.

\FloatBarrier
\subsection{Funnel accuracy at scale}
\label{sec:eval-scale}

\begin{table}[!htbp]
\centering
\caption{Recall against query latency of the coarse tier, for the shipped one-bit code and an affine
three-bit code at three shortlist widths, over 64 queries on 500{,}000 rows of seeded vectors. Recall
is the share of an exact fp32 top 10 that the funnel returns in its own top 10, and $C = 7{,}680$ is
the shipped width.}
\label{tab:scale}
\small
\setlength{\tabcolsep}{5pt}
\begin{tabular}{lrrrrrrr}
\toprule
& & & \multicolumn{5}{c}{query p50 (ms)} \\
\cmidrule(lr){4-8}
Tier & $C$ & recall@10 & M3 Ultra & M4 Pro & M4 & M3 Pro & M2 \\
\midrule
one bit  & 3{,}840  & 92.8 & 5.4 & 5.5 & 5.8 & 5.4 & 6.9 \\
one bit  & 7{,}680  & 95.8 & 7.0 & 6.6 & 7.7 & 7.3 & 9.5 \\
one bit  & 15{,}360 & 97.3 & 9.5 & 8.8 & 11.0 & 10.1 & 13.2 \\
three bit & 3{,}840  & 97.8 & 5.1 & 5.4 & 6.6 & 5.8 & 7.9 \\
three bit & 7{,}680  & 97.8 & 6.8 & 6.8 & 8.3 & 7.7 & 10.4 \\
three bit & 15{,}360 & 97.8 & 9.3 & 9.0 & 11.7 & 10.5 & 14.0 \\
\bottomrule
\end{tabular}
\end{table}

\Cref{tab:scale} gives the recall the funnel keeps relative to an exact scan. It is identical on all
five machines, since the vectors are seeded and the exact top 10 is computed from the same rows. At
every shortlist the wider code is the more accurate scan, and it is already saturated at the
narrowest one; the one-bit tier needs four times the shortlist to come within half a point of it.
What the narrow tier saves is memory:
at $d/8$ bytes a row it holds a sixteenth of what the exact vectors hold and under a third of what an
affine code at three bits holds. The accuracy it gives up is recovered by the shortlist: at the
shipped width, twice the narrowest, it trails the three-bit code by two points.

The shortlist is a fixed 7{,}680 candidates, so it covers a smaller share of a larger index: 1.5\%
of the 500{,}000 rows above and 0.1\% of eight million. \Cref{tab:recallscale} repeats the shipped
point as the index grows, and the funnel keeps 88.3\% of the exact top 10 at eight million rows.

\begin{table}[!htbp]
\centering
\caption{The shipped point, one bit at $C = 7{,}680$, as the index grows: recall@10 against an exact
fp32 top 10 over 64 queries of seeded vectors, and the p50 search time on the M3 Ultra.}
\label{tab:recallscale}
\small
\setlength{\tabcolsep}{5pt}
\begin{tabular}{rrrr}
\toprule
Rows & $C$ / rows & recall@10 & search p50 (ms) \\
\midrule
1{,}000{,}000 & 0.77\% & 95.3 & 7.8 \\
2{,}000{,}000 & 0.38\% & 94.1 & 8.7 \\
4{,}000{,}000 & 0.19\% & 89.8 & 9.9 \\
8{,}000{,}000 & 0.10\% & 88.3 & 10.6 \\
\bottomrule
\end{tabular}
\end{table}

\FloatBarrier
\subsection{Exact shortlist selection}
\label{sec:eval-select}

Once the coarse scan has produced its scores, the shortlist has to be selected from them. The
primitive the framework offers routes internally to a full sort, so asking for a few thousand rows
out of millions sorts millions. In isolation it costs 0.7 and 1.9\,ms at one and four million
rows on the M3 Ultra, and 2.9 and 12.5\,ms at those sizes on the M2.

We reduce that cost without changing the result. The scores are cut into tiles of 32 rows and the
$C$ tiles with the highest maxima are kept. This is exact, because any row in the true top $C$ lies
in a tile whose maximum reaches the $C$-th largest tile maximum, and it costs one selection over
$N/32$ values and one over $32C$, against a sort over $N$. At four million rows it returns the same
indices as the primitive 3.6 to 9.5$\times$ faster. The two-level form is faster only well above the
shortlist width, so below $128\,C$ rows, 983{,}040 at the shipped width, the single call is kept.

\FloatBarrier
\subsection{Funnel crossover}
\label{sec:eval-cross}

\begin{table}[!htbp]
\centering
\caption{The p50 latency of the exact scan divided by that of the funnel, over 40 queries at each
size. A value above one means the funnel is faster, and bold marks where each machine first passes
one. Both arms use the same seeded vectors, and a size runs only where both fit: 500{,}000 chunks need
16\,GiB of memory, one million 24 and two million 32.}
\label{tab:cross}
\footnotesize
\setlength{\tabcolsep}{4pt}
\begin{tabular}{rrrrrr}
\toprule
Chunks & \shortstack{M3 Ultra\\2025} & \shortstack{M4 Pro\\2024} & \shortstack{M4\\2024}
       & \shortstack{M3 Pro\\2023} & \shortstack{M2\\2022} \\
\midrule
125{,}000 & 0.51 & 0.62 & 0.73 & 0.67 & 0.66 \\
250{,}000 & 0.60 & 0.73 & 1.00 & 0.86 & 0.93 \\
500{,}000 & 0.60 & 0.90 & \textbf{1.32} & \textbf{1.30} & \textbf{1.40} \\
1{,}000{,}000 & 0.64 & \textbf{1.26} & --- & --- & --- \\
2{,}000{,}000 & \textbf{1.11} & 2.18 & --- & --- & --- \\
\bottomrule
\end{tabular}
\end{table}

\Cref{tab:cross} shows where each machine crosses over. The M3 Ultra, with eight times the GPU cores
of the narrowest machine, crosses last, because its exact scan is already fast enough that the
replica has little time to save. The shipped policy adopts the replica past 250{,}000 rows below
sixteen GPU cores, past 500{,}000 below thirty-two and past one million above that. At those thresholds it adds under a
millisecond to a query and holds a sixteenth of the memory of the exact vectors. On the CPU, where
the store scans with the encoder on the Neural Engine, the replica is level with the
exact scan at 125{,}000 rows on the M3 Ultra and 3.3$\times$ faster at two million, so that
mode adopts it past 125{,}000 rows. On every machine
the replica is also adopted once the exact matrix would pass its memory budget, near one million
rows for \modelnano{} on a Mac with 16\,GB or more.

\FloatBarrier
\subsection{The cost of a deletion}
\label{sec:eval-delete}

\begin{table}[!htbp]
\centering
\caption{The cost of removing one file on the M3 Ultra at two index sizes, with 40 deletions spread
across the row space and one query after each. The slope is the cost at 500k rows over the cost at
125k, so one means the cost does not grow with the index. The compacting arm compacts the resident matrix on every deletion, and the shipped arm marks
the rows dead.}
\label{tab:delete}
\small
\setlength{\tabcolsep}{5pt}
\begin{tabular}{lrrr}
\toprule
& \multicolumn{2}{c}{delete p50 (ms)} & \\
\cmidrule(lr){2-3}
Arm & 125k rows & 500k rows & slope \\
\midrule
compacting  & 98.0 & 453 & 4.62 \\
marked dead & 0.22 & 0.32 & 1.43 \\
\bottomrule
\end{tabular}
\end{table}

\Cref{tab:delete} compares the two. Compacting on every deletion costs 4.6 times as much at four
times the index, close to the growth the row copy implies. Marking the rows dead keeps a deletion
at 0.22 and 0.32\,ms at the two sizes.

\FloatBarrier
\subsection{Memory under one budget}
\label{sec:eval-cap}

\Cref{tab:memory} gives the peak footprint of each phase on every machine.

\begin{table}[!htbp]
\centering
\caption{Peak footprint of the whole process, sampled every 50\,ms, against what the setting
allows, which is the resident set at rest plus the 1\,GB working floor and the 1\,GB of headroom. The
process indexes every kind of file into a copy of the one-million-chunk store while searching it
every 100\,ms, then deletes 50{,}000 files and compacts the store.}
\label{tab:memory}
\footnotesize
\setlength{\tabcolsep}{4pt}
\begin{tabular}{lrrrrr}
\toprule
& \shortstack{M3 Ultra\\2025} & \shortstack{M4 Pro\\2024} & \shortstack{M4\\2024}
& \shortstack{M3 Pro\\2023} & \shortstack{M2\\2022} \\
\midrule
At rest, model and index loaded (MB) & 11{,}311 & 6{,}049 & 5{,}775 & 5{,}784 & 4{,}380 \\
Indexing every kind, searching (MB)   & 13{,}666 & 8{,}139 & 7{,}429 & 7{,}613 & 5{,}760 \\
Deleting and compacting (MB)          & 12{,}419 & 6{,}925 & 6{,}504 & 6{,}554 & 5{,}270 \\
Allowed: rest, floor, headroom (MB)   & 13{,}289 & 8{,}027 & 7{,}753 & 7{,}762 & 6{,}357 \\
Peak over what is allowed (MB)        & 377 & 112 & $-$323 & $-$149 & $-$598 \\
\bottomrule
\end{tabular}
\end{table}

\end{document}